\documentclass[11pt]{article}
\usepackage{graphicx}

\newcommand{\BABARPubYear}    {03}

\newcommand{\BABARConfNumber} {11}
\newcommand{\SLACPubNumber}   {9684}

\RequirePackage{xspace}

\usepackage{relsize}
\def\babar{\mbox{\slshape B\kern-0.1em{\smaller A}\kern-0.1em
    B\kern-0.1em{\smaller A\kern-0.2em R}}}

\def\epem       {\ensuremath{e^+e^-}\xspace}

\def\tautau     {\ensuremath{\tau^+\tau^-}\xspace}

\def\gaga  {\ensuremath{\gamma\gamma}\xspace}  

\def\qqbar {\ensuremath{q\overline q}\xspace}

\def\uubar {\ensuremath{u\overline u}\xspace}

\def\s     {\ensuremath{s}\xspace}
\def\sbar  {\ensuremath{\overline s}\xspace}

\def\pip   {\ensuremath{\pi^+}\xspace}
\def\pim   {\ensuremath{\pi^-}\xspace}

\def\pipm  {\ensuremath{\pi^\pm}\xspace}

\def\kaon  {\ensuremath{K}\xspace}
\def\Kbar  {\kern 0.2em\overline{\kern -0.2em K}{}\xspace}

\def\Kz    {\ensuremath{K^0}\xspace}
\def\Kzb   {\ensuremath{\Kbar^0}\xspace}
\def\KzKzb {\ensuremath{\Kz \kern -0.16em \Kzb}\xspace}
\def\Kp    {\ensuremath{K^+}\xspace}
\def\Km    {\ensuremath{K^-}\xspace}
\def\Kpm   {\ensuremath{K^\pm}\xspace}

\def\KpKm  {\ensuremath{\Kp \kern -0.16em \Km}\xspace}
\def\KS    {\ensuremath{K^0_{\scriptscriptstyle S}}\xspace}

\def\Dbar    {\kern 0.2em\overline{\kern -0.2em D}{}\xspace}

\def\Dz      {\ensuremath{D^0}\xspace}
\def\Dzb     {\ensuremath{\Dbar^0}\xspace}
\def\DzDzb   {\ensuremath{\Dz {\kern -0.16em \Dzb}}\xspace}
\def\Dp      {\ensuremath{D^+}\xspace}
\def\Dm      {\ensuremath{D^-}\xspace}

\def\DpDm    {\ensuremath{\Dp {\kern -0.16em \Dm}}\xspace}
\def\Dstar   {\ensuremath{D^*}\xspace}

\def\B       {\ensuremath{B}\xspace}
\def\Bbar    {\kern 0.18em\overline{\kern -0.18em B}{}\xspace}

\def\BB      {\ensuremath{B\Bbar}\xspace} 
\def\Bz      {\ensuremath{B^0}\xspace}
\def\Bzb     {\ensuremath{\Bbar^0}\xspace}
\def\BzBzb   {\ensuremath{\Bz {\kern -0.16em \Bzb}}\xspace}
\def\Bu      {\ensuremath{B^+}\xspace}
\def\Bub     {\ensuremath{B^-}\xspace}

\def\Bpm     {\ensuremath{B^\pm}\xspace}

\def\BpBm    {\ensuremath{\Bu {\kern -0.16em \Bub}}\xspace}

\mathchardef\Upsilon="7107
\def\Y#1S{\ensuremath{\Upsilon{(#1S)}}\xspace}

\def\FourS {\Y4S}

\mathchardef\Deltares="7101
\mathchardef\Xi="7104
\mathchardef\Lambda="7103
\mathchardef\Sigma="7106
\mathchardef\Omega="710A

\def\Deltabar{\kern 0.25em\overline{\kern -0.25em \Deltares}{}\xspace}
\def\Lbar{\kern 0.2em\overline{\kern -0.2em\Lambda\kern 0.05em}\kern-0.05em{}\xspace}
\def\Sigbar{\kern 0.2em\overline{\kern -0.2em \Sigma}{}\xspace}
\def\Xibar{\kern 0.2em\overline{\kern -0.2em \Xi}{}\xspace}
\def\Obar{\kern 0.2em\overline{\kern -0.2em \Omega}{}\xspace}
\def\Nbar{\kern 0.2em\overline{\kern -0.2em N}{}\xspace}
\def\Xb{\kern 0.2em\overline{\kern -0.2em X}{}\xspace}

\def\BR         {{\ensuremath{\cal B}\xspace}}

\def\mes        {\mbox{$m_{\rm ES}$}\xspace}

\def\DeltaE     {\mbox{$\Delta E$}\xspace}

\newcommand{\tev}{\ensuremath{\mathrm{\,Te\kern -0.1em V}}\xspace}
\newcommand{\gev}{\ensuremath{\mathrm{\,Ge\kern -0.1em V}}\xspace}
\newcommand{\mev}{\ensuremath{\mathrm{\,Me\kern -0.1em V}}\xspace}
\newcommand{\kev}{\ensuremath{\mathrm{\,ke\kern -0.1em V}}\xspace}
\newcommand{\ev}{\ensuremath{\mathrm{\,e\kern -0.1em V}}\xspace}
\newcommand{\gevc}{\ensuremath{{\mathrm{\,Ge\kern -0.1em V\!/}c}}\xspace}
\newcommand{\mevc}{\ensuremath{{\mathrm{\,Me\kern -0.1em V\!/}c}}\xspace}
\newcommand{\gevcc}{\ensuremath{{\mathrm{\,Ge\kern -0.1em V\!/}c^2}}\xspace}
\newcommand{\mevcc}{\ensuremath{{\mathrm{\,Me\kern -0.1em V\!/}c^2}}\xspace}

\def\cm   {\ensuremath{\rm \,cm}\xspace}

\def\invfb   {\ensuremath{\mbox{\,fb}^{-1}}\xspace}

\def\mus  {\ensuremath{\rm \,\mus}\xspace}

\def\mus        {\ensuremath{\,\mu{\rm s}}\xspace}    

\def\to                 {\ensuremath{\rightarrow}\xspace}

\def\pep2{PEP-II}

\newcommand{\dedx}{\ensuremath{\mathrm{d}\hspace{-0.1em}E/\mathrm{d}x}\xspace}

\def\gsim{{~\raise.15em\hbox{$>$}\kern-.85em
          \lower.35em\hbox{$\sim$}~}\xspace}
\def\lsim{{~\raise.15em\hbox{$<$}\kern-.85em
          \lower.35em\hbox{$\sim$}~}\xspace}

\def\CP                {\ensuremath{C\!P}\xspace}

\xspace

\def\jetset74   {\mbox{\tt Jetset \hspace{-0.5em}7.\hspace{-0.2em}4}\xspace}

\def\vphi{\ensuremath{\phi}\xspace}

\def\hpm {\ensuremath{h^\pm}\xspace}

\def\phiK{\ensuremath{\B \to \vphi \kaon}\xspace}
\def\phipi{\ensuremath{\B \to \vphi \pi}\xspace}
\def\phiKpm{\ensuremath{\Bpm \to \vphi \Kpm}\xspace}
\def\phiKz{\ensuremath{\Bz \to \vphi \Kz}\xspace}
\def\phiKs{\ensuremath{\Bz \to \vphi \KS}\xspace}
\def\phih{\ensuremath{B \to \vphi h}\xspace}
\def\phihpm{\ensuremath{\Bpm \to \vphi \hpm}\xspace}
\def\phipipm{\ensuremath{\Bpm \to \vphi \pipm}\xspace}

\def\piDzbarKpi{\ensuremath{\Bu \to \pip \Dzb \ (\Dzb \to \Kp \pim) }\xspace}

\def\piDmKzpi{\ensuremath{\Bz \to \pip \Dm \ (\Dm \to \Kz \pim) }\xspace}

\def\phiKK{\ensuremath{\vphi \to \Kp \Km}\xspace}

\def\Kspipi{\ensuremath{\KS \to \pip \pim}\xspace}

\def\DzKpi{\ensuremath{\Dz \to \Km \pip}\xspace}

\def\btosss{\ensuremath{b\to\s\bar{s}s}\xspace}

\def\abscosThT{\ensuremath{|\cos{\theta_T}|}\xspace}

\def\cosThH{\ensuremath{\cos{\theta_H}}\xspace}

\def\mKK{\mbox{$m_{KK}$}\xspace}
\def\ACP{\mbox{${\cal A}_{CP}$}\xspace}

\long\def\inst#1{\par\nobreak\kern 4pt\nobreak
    {\it #1}\par\vskip 10pt plus 3pt minus 3pt}

\def\low#1pad{\raisebox{-#1pt}{\rule{0pt}{8pt}}}
\def\high#1pad{\raisebox{#1pt}{\rule{0pt}{8pt}}}

\def\babar{\mbox{\slshape B\kern-0.37em{\relsize{-2} A}\kern-0.04em
    B\kern-0.37em{\relsize{-2} A\kern-0.04em R}}\xspace}

\begin{document}
{\pagestyle{empty}


\begin{flushright}
\babar-CONF-\BABARPubYear/\BABARConfNumber \\
SLAC-PUB-\SLACPubNumber \\
March 2003 \\
\end{flushright}

\par\vskip 2.5cm

\begin{center}
\Large \bf \boldmath Branching Fractions in \phih and \\ Search for Direct \CP Violation in \phiKpm
\end{center}
\bigskip

\begin{center}
\large The \babar\ Collaboration\\
\mbox{ }\\
March 20, 2003
\end{center}
\bigskip \bigskip

\begin{center}
\large \bf Abstract
\end{center}

We present preliminary measurements of branching fractions of the
\btosss penguin-dominated decays \phiKpm and \phiKz in a sample of approximately 89 million \BB pairs
collected by the \babar detector at the \pep2\ asymmetric-energy \emph{B}-meson Factory at SLAC. 
We determine $\BR(\phiKpm) = (10.0^{+0.9}_{-0.8}\,\mbox{(stat.)} \pm 0.5\,\mbox{(syst.)}) \times 10^{-6}$ and
$\BR(\phiKz) = (7.6^{+1.3}_{-1.2}\,\mbox{(stat.)} \pm 0.5\,\mbox{(syst.)}) \times 10^{-6}$. Additionally, we measure the
charge asymmetry $\ACP(\phiKpm) = 0.039 \pm 0.086\,\mbox{(stat.)} \pm 0.011\,\mbox{(syst.)}$ and 
set an upper limit on the CKM-- and color-suppressed
decay \phipipm, $\BR(\phipipm) < 0.41 \times 10^{-6}$ (90\% CL).
\vfill

\begin{center}
Presented at the XXXVIII$^{th}$ Rencontres de Moriond on\\
Electroweak Interactions and Unified Theories, \\
3/15--3/22/2003, Les Arcs, Savoie, France
\end{center}

\vspace{1.0cm}
\begin{center}
{\em Stanford Linear Accelerator Center, Stanford University, 
Stanford, CA 94309} \\ \vspace{0.1cm}\hrule\vspace{0.1cm}
Work supported in part by Department of Energy contract DE--AC03--76SF00515.
\end{center}

\newpage
} 

\begin{center}
\small

The \babar\ Collaboration,
\bigskip

%
B.~Aubert,
R.~Barate,
D.~Boutigny,
J.-M.~Gaillard,
A.~Hicheur,
Y.~Karyotakis,
J.~P.~Lees,
P.~Robbe,
V.~Tisserand,
A.~Zghiche
\inst{Laboratoire de Physique des Particules, F-74941 Annecy-le-Vieux, France }
% ...

\end{center}\newpage


\section{Introduction}
\label{sec:Introduction}

\begin{figure}[!htbp]
\begin{center} 
  \begin{tabular}{c c}
    \includegraphics[width=3.9in]{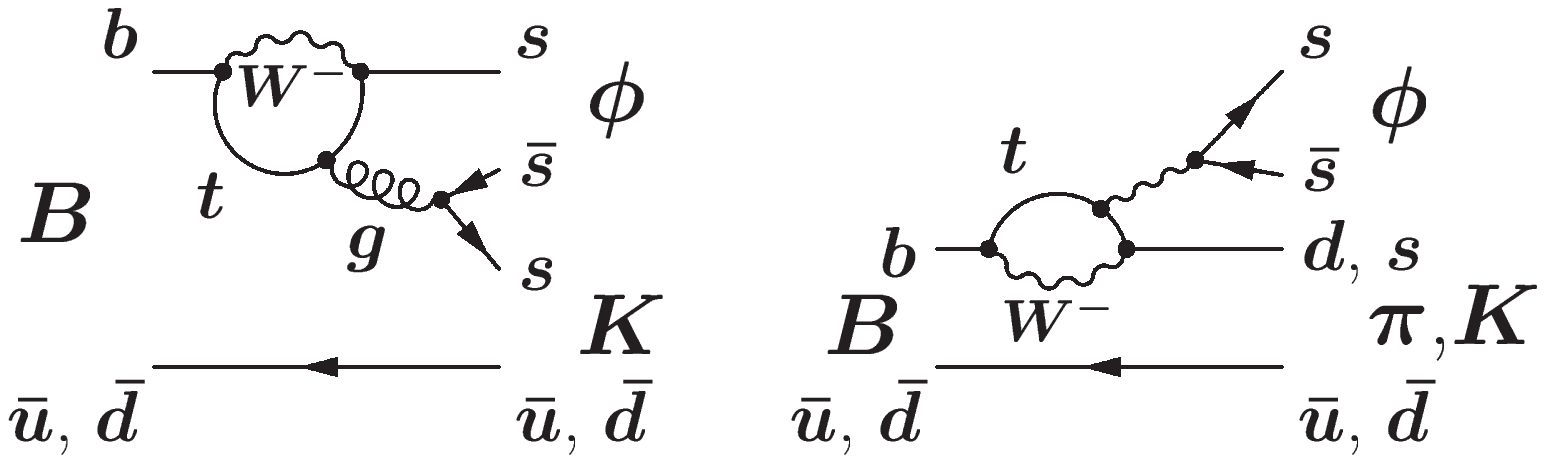} &
    \raisebox{0.32in}{\includegraphics[width=2.3in]{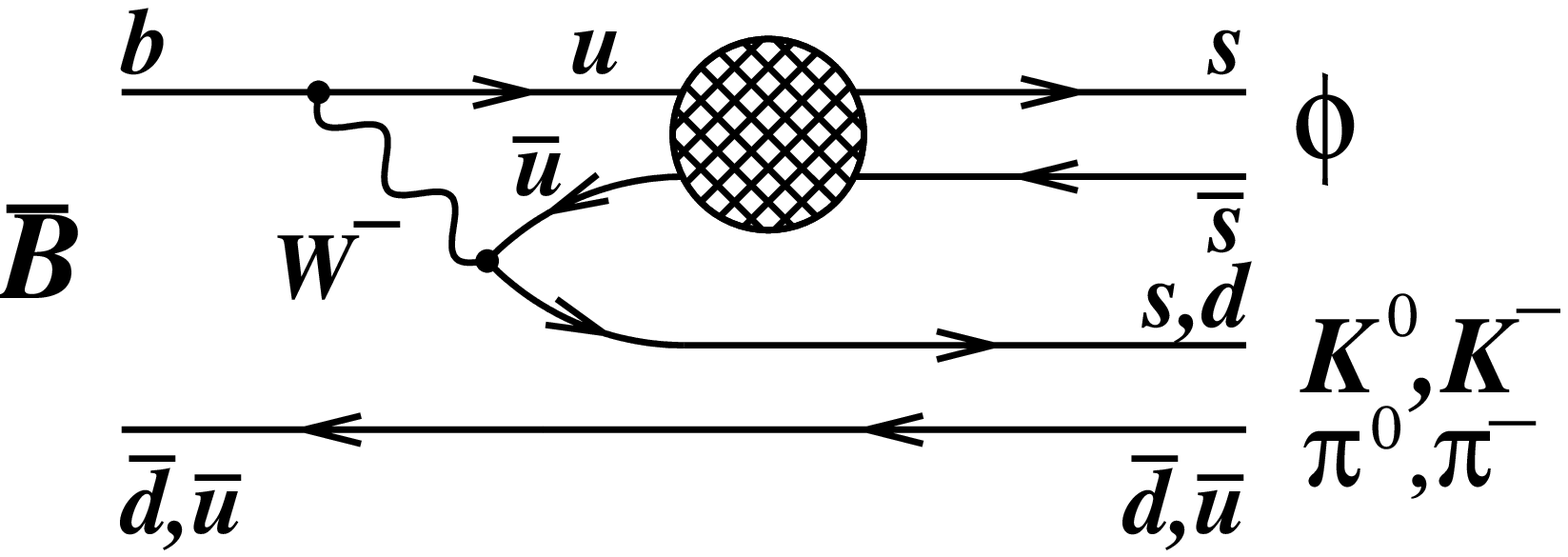}} \\
  \end{tabular}
  \put(-105,-33){\small \bfseries (c)}
  \put(-270,-33){\small \bfseries (b)}
  \put(-410,-33){\small \bfseries (a)}
  \put(-32,3){\large \bfseries --}
  \put(-200,-3){\huge \bfseries --}
  \put(-314,-5){\huge \bfseries --}
  \put(-356,-5){\huge \bfseries --}
  \put(-459,15){\huge \bfseries --}
  
  \vspace*{-0.2in}
  \caption{\small Examples of penguin diagrams for \emph{(a)}~\phiK, \emph{(b)}~\phiK and \phipi; \emph{(c)}~the rescattering diagram for \phiK and \phipi.
    The unlabled spin-1 propagator in diagram (b) should be interpreted either as a hard gluon (with two or more additional soft gluons that 
    are not shown), or as a $\gamma$ or a $Z$. The tree part of diagram (c) could lead directly to a \phih final state via the small, poorly
    understood \uubar component of the \vphi resonance  
   }
  \label{fig:Penguins}
\end{center}
\end{figure}

$B$-meson decays with a \vphi in the final state present a special interest because 
they are dominated by the $b \to s(d)\bar{s}s$ gluonic penguins, 
possibly with a significant contribution from electroweak penguins (Figs.~\ref{fig:Penguins}a,~\ref{fig:Penguins}b), while other Standard Model (SM)
contributions are strongly suppressed~\cite{ref:theory1}. Since contributions of diagrams with a $c$\/ or a $u$\/ quark in 
the loop are small, all dominant SM decay amplitudes have the same weak phase, leading to a very small ($\sim 1\%$) predicted
value of direct $CP$\/ asymmetry $\ACP$
in \phiKpm. However, many models of new physics introduce new heavy particles, with
new couplings, that would contribute to these decays, potentially making \ACP quite large~\cite{ref:newphys}. 
The amounts of \CP and flavor violation observed in these decays can therefore be used
to constrain the parameters of models of new physics.

\vspace*{-0.1in}
Recent preliminary results from \babar and Belle on the time-dependent \CP asymmetry in the decay 
\phiKs~\cite{ref:phiKSexp_BaBar,ref:phiKSexp_Belle} have raised questions about the 
rescattering contribution to the \phiKz decay amplitude (Fig.~\ref{fig:Penguins}c).
While this cannot be computed \emph{a priori}, simply from the weak couplings it will be larger in
\phipi than in \phiK by a factor of roughly $\cot(\theta_\mathrm{C}) \approx 4.4$, 
where $\theta_\mathrm{C}$ is the Cabibbo angle~\cite{Grossman:1997gr}.
By searching for \phipipm, our analysis can 
constrain the magnitude of the rescattering contribution to \phiKz.

Additional reasons to be interested in a detailed study of the $b \to s(d)\bar{s}s$ processes include
their sensitivity to QCD dynamics~\cite{Cheng:2000hv, Chen:2001pr} 
and to the poorly measured Cabibbo--Kobayashi--Maskawa matrix element $V_{ts}$. 

The decays \phiKpm and \phiKz
have previously been observed by CLEO~\cite{Briere:2001ue}, \babar~\cite{Aubert:2001zd}, and Belle~\cite{Bozek:2001xd}.
The significantly increased size of the \babar data set and an improved analysis technique
allow us to achieve a substantial reduction of both the statistical and the systematic errors on the branching fractions
of the two decays.
The analysis is based on a multivariate maximum-likelihood fit;
the yields for the decay modes \phiKpm and \phipipm are obtained simultaneously.
A blind analysis technique
is used to avoid the potential for an experimenter-induced bias: the signal region is hidden 
until all significant details of the analysis are finalized. The determination of systematic errors is completed subsequently. 

\section{The \babar Detector and Data Set}
\label{sec:babar}

The data were collected with the \babar detector~\cite{ref:babar} in 1999--2002
at the \pep2\ asymmetric-energy \epem collider~\cite{ref:pep}
located at the Stanford Linear Accelerator Center.
An integrated
luminosity of about 82~\invfb was recorded at the peak of the \FourS resonance, 
corresponding to $88.9\pm1.0$ million \BB pairs.

The asymmetric beam configuration
provides a boost to the \FourS in the laboratory frame ($\beta\gamma\approx 0.56$),
increasing the maximum momentum of the $B$-meson decay products
to $4.4 \gevc$.
Charged particles are detected and their momenta measured
by a combination of a silicon vertex tracker (SVT), consisting 
of five double-sided layers, and a 40-layer central drift chamber 
(DCH), both operating in a 1.5~T solenoidal magnetic field. 
The tracking system covers 92\% of the solid angle
in the center-of-mass (CM) frame.
The track-finding efficiency is, on average, ($98\pm1$)\% for momenta
above $0.2 \gevc$ and polar angles greater than 0.5~rad. 
Photons are detected by a CsI electromagnetic calorimeter (EMC), which
provides excellent angular and energy resolution with high efficiency for 
energies above 20~\mev.

Charged-particle identification is provided by the average 
energy loss (\dedx) in the two tracking devices and
by the novel internally reflecting ring-imaging 
Cherenkov detector (DIRC) covering the central region. 
A $\pi/K$ separation of better than $4\sigma$ is 
achieved for tracks with momenta below $3 \gevc$, decreasing to 
$2.5\sigma$ for the highest momenta arising from $B$-meson decays. 
Electrons are identified with the use
of the tracking system and the EMC.

\section{Event Selection}
\label{sec:EventSelection}

Hadronic events are selected on the basis of track multiplicity and 
event topology. $B$-meson 
candidates are fully reconstructed from their charged
decay products: \phiKK and $\Kz\to\Kspipi$.
Charged tracks that are $B$\/ or $\vphi$ daughters are required to originate 
from the interaction point (within 
10~\cm along the beam direction and 1.5~\cm in the transverse plane),
have at least 12 DCH hits 
and a minimum transverse momentum of $0.1 \gevc$. 
Looser criteria are applied to tracks used to reconstruct \Kspipi candidates
in order to allow for displaced \KS decay vertices. We suppress $\epem \to \tautau$ 
and $\epem \to \epem \gaga$ backgrounds by rejecting events with fewer than
5 tracks.

\begin{figure}[tb]
\begin{center}
\begin{tabular}{c c}
  \includegraphics[height=2.40in]{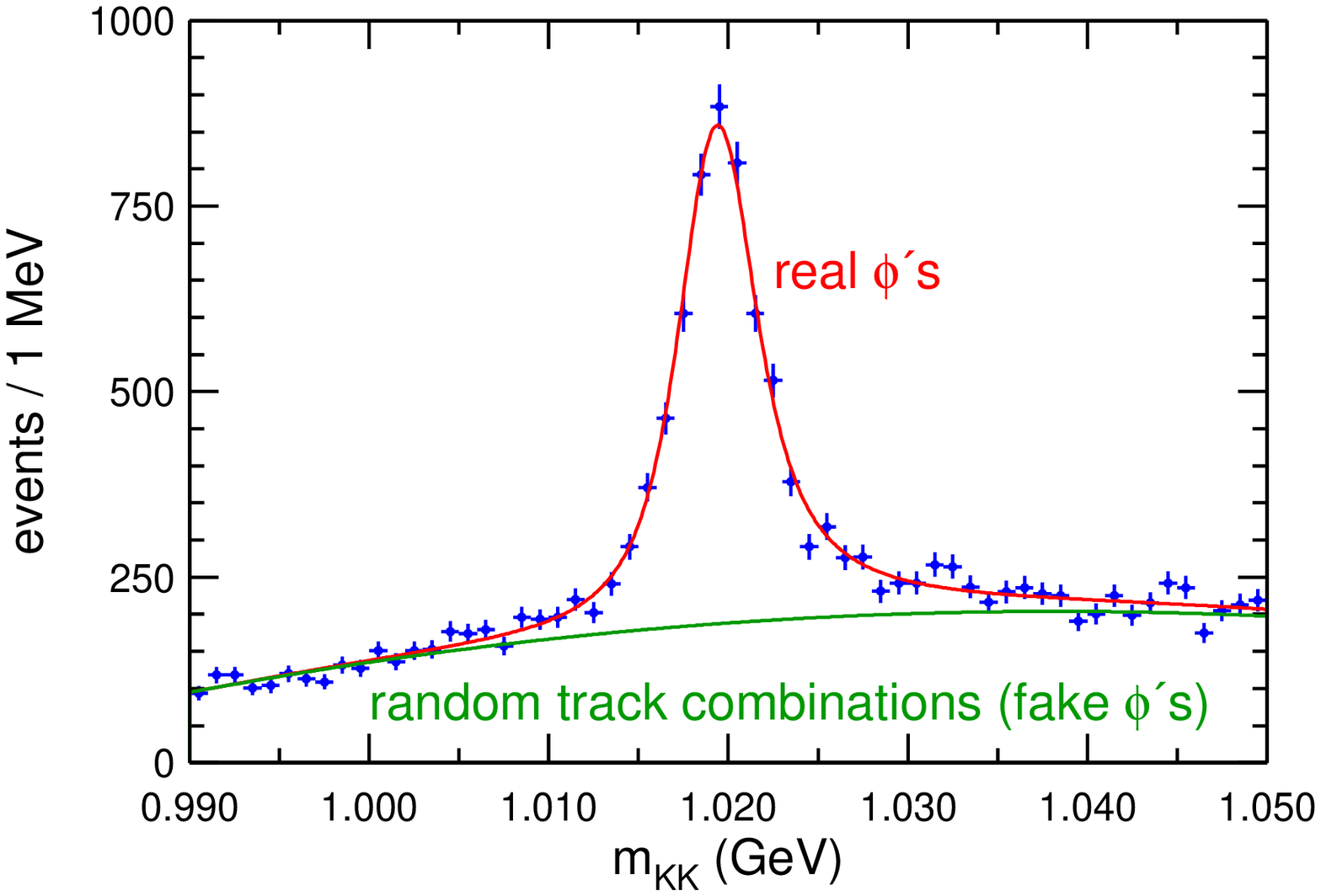} 
     & \includegraphics[height=2.40in]{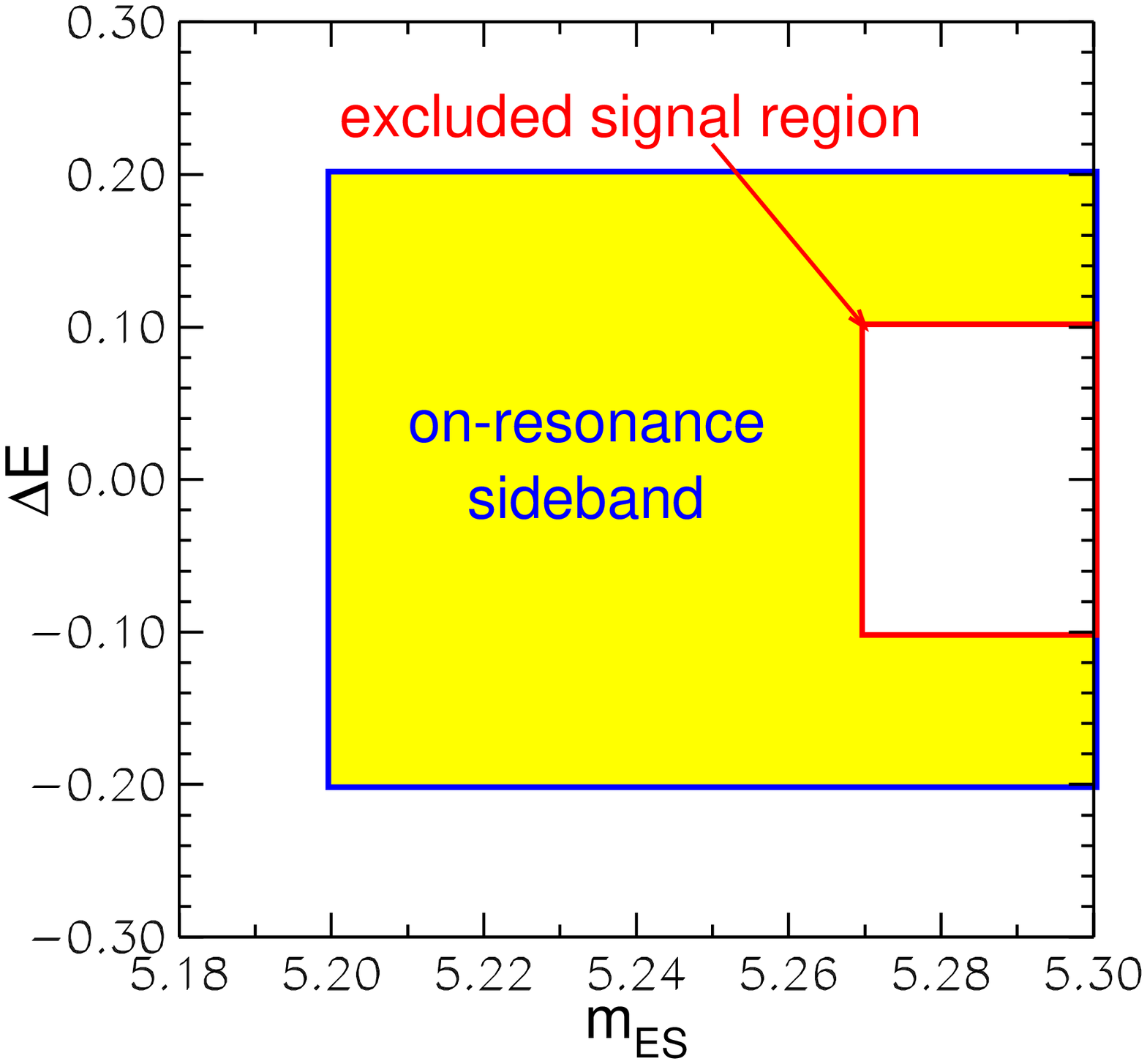} \\  
\end{tabular} 
  \put(-419,75){\bfseries (a)}
  \put(-160,75){\bfseries (b)}
  \put(-419,60){\small \phihpm sideband}
  \caption{\normalsize \emph{(a)} \mKK invariant mass distribution in the \phihpm on-resonance sideband; 
   \emph{(b)} definition of the on-resonance sideband in \mes and \DeltaE}
  \label{fig:phiMass}
\end{center}
\end{figure}

Pairs of oppositely-charged tracks that 
are required to originate from a common
vertex are combined to form the \vphi and \KS candidates.
A clean sample of \KS candidates is obtained with requirements
on the two-pion invariant mass 
($|M_{\pip\pim} - m_{K^0}|< 12 \mevcc$), 
the angle $\alpha$ between the reconstructed flight 
and momentum directions ($\cos\alpha >$ 0.995) 
and the flight-length significance ($\ell/\sigma_\ell >$ 3).
For \phiKK, the invariant mass of the $\Kp\Km$ pair is
required to lie within the $[0.99, 1.05] \gevcc$ range (Fig.~\ref{fig:phiMass}).

Tracks used to reconstruct the \phiKK decay are distinguished from 
pion and proton tracks via a relatively loose requirement on a 
likelihood ratio that includes, for momenta below $0.7 \gevc$, 
\dedx information from the SVT and DCH and, for higher
momenta, the Cherenkov angle and the number of photons
as measured by the DIRC. In addition, these tracks are required to 
pass electron and proton vetoes.
Determination of the flavor of the high-momentum \hpm track in \phihpm decays is
provided mostly by Cherenkov-angle residuals, normalized to their uncertainties, 
which are computed for the pion and the kaon
hypotheses and are used in the maximum-likelihood fit. During event selection, \hpm candidates
are required to have Cherenkov angles consistent within $\pm 4 \sigma$ with either of the two hypotheses; they 
are also required to pass an electron veto.

We identify $B$-meson candidates kinematically
using two nearly independent variables \cite{ref:babar}: 
the beam-energy--substituted mass
$\mes = \sqrt{ ((\frac{s}{2} + {\vec{p}}_{\Upsilon} \cdot {\vec{p}}_{B} )^{2})/({E^{2}_{\Upsilon}}) - {{\vec{p}}_{B}}^{\,2}}$, 
which is computed in the laboratory frame
and is independent of the mass hypotheses assigned to the $B$-candidate daughters,
and
the Lorentz-invariant missing energy
$\Delta E = (q_{\Upsilon} \cdot q_{B}/\sqrt{s}) - \sqrt{s}$.
Here $q_{\Upsilon}$\/ and $q_{B}$ are four-momenta of the \Y4S and the $B$\/ 
candidate, $s \equiv (q_{\Upsilon})^{2}$ is the square of the center-of-mass energy, ${\vec{p}}_{\Upsilon}$ 
and ${\vec{p}}_{B}$ are the three-momenta of the \Y4S\/ and the $B$\/ in the laboratory frame, 
and ${E_{\Upsilon}} \equiv q^0_{\Upsilon}$ is the energy of the \Y4S in the laboratory frame.
For signal events, \DeltaE peaks at zero and
\mes peaks at the $B$\/ mass. 
Our selection requires
$|\DeltaE|<0.2 \gev$ and $\mes > 5.2 \gevcc$. Being dependent on the mass hypotheses
assigned to the $B$\/ decay products, \DeltaE provides additional momentum-dependent $\pi/K$
separation in the maximum-likelihood fit for \phihpm branching fractions.

Detailed Monte Carlo studies demonstrate that backgrounds from other
$B$\/ decays are negligible.
Backgrounds are dominated by random 
combinations of tracks produced in the quark-antiquark (\qqbar) 
continuum.  
This background is distinguished by its jet-like structure---as 
opposed to the nearly spherical decay of the \FourS.
We have considered a variety of CM event-shape variables that 
exploit this difference.

One such variable is \abscosThT, where $\theta_T$ is the angle 
between the thrust axis of the $B$\/ candidate and the thrust axis 
of the rest of the event, where the thrust axis $\vec{A}$\/ is defined as the
unit vector that maximizes the thrust
$T= \max \left({\sum_{i=1}^{N}|\vec{A}\cdot\vec{p_i}|}/{\sum_{i=1}^{N}\sqrt{|\vec{p_i}|^2}}\right)$.
Since $B$'s are non-relativistic in the \Y4S rest frame ($\beta \approx 0.06$), 
the \abscosThT distribution for true $B$\/ candidates is very well described by a nearly flat first-degree polynomial;
on the other hand, the \abscosThT distribution for $B$\/ candidates 
found in the $\epem\to\qqbar$ continuum is sharply peaked at $+1$. We apply the cut $\abscosThT < 0.9$ throughout our analysis.

Other quantities that characterize the event shape are the
$B$\/ polar angle $\theta_B$ and the angle $\theta_{q\bar{q}}$ 
of the $B$-candidate thrust axis, both defined with respect to the beam axis,
as well as the zeroth and the second Legendre moments
of the rest of the tracks and neutrals, ${\cal L}_n = \sum p_i \times L_n(\theta_i)$, 
$n=0$, 2, computed relative to the $B$-candidate thrust axis.
For \Y4S decays into two pseudoscalar $B$\/ mesons, 
the $\theta_B$ distribution has a $\sin^2\theta_B$ 
dependence, whereas the jets from continuum events lead to a 
uniform distribution in $\cos\theta_B$.
In $\theta_{q\bar{q}}$, the continuum jets give rise to a 
$(1+\cos^2\theta_{q\bar{q}})$ distribution, 
while the thrust direction of true $B$\/ decays is random.
We further suppress the background by forming an optimized
linear combination (Fisher discriminant~\cite{Fisher:et}) of the four variables:
$|\cos\theta_B|$, $|\cos\theta_{q\bar{q}}|$, ${\cal L}_0$ and ${\cal L}_2$.

\section{Maximum Likelihood Fit}
\label{sec:MLfit}

We use an unbinned extended maximum-likelihood (ML) fit to extract
signal yields and charge asymmetries simultaneously.
The extended likelihood for a sample of $N$\/ events is 
\begin{equation}
{\cal L} = \exp\left(-\sum_{i,k}^{} n_{ik}\right)\, \prod_{j=1}^N 
\left(\sum_{i,k}  n_{ik}\, 
{\cal P}_{ik}(\vec{x}_j;\vec{\alpha})\right) ,
\label{eq:likel}
\end{equation}
where ${\cal P}_{ik}(\vec{x}_j;\vec{\alpha})$ is the probability density function (PDF) for measured variables $\vec{x}_j$ of an 
event $j$ in category $i$ and flavor state $k$, and $n_{ik}$ are 
the yields extracted from the fit.
The fixed parameters $\vec{\alpha}$ describe the expected 
distributions of measured variables in each category and 
flavor state.
The PDFs are non-zero only for the correct final 
state flavor ($k = 1$ for $\Bbar\rightarrow\bar f$ and 
$k = 2$ for $B\rightarrow f$).
In the simplest case, there are two categories, 
signal and background ($i=1,2$).
The decays with a charged primary daughter 
\phihpm 
($h= \pi$ or $K$)
are fitted simultaneously with two signal 
($i=1$ for \phiKpm and 
$i=2$ for \phipipm)
and two corresponding background ($i=3,4$) categories. 

We define the event yields $n_{ik}$ in each category in terms of 
the asymmetry ${\cal A}_i$ and the total event yield $n_{i}$:
$n_{i1} = n_{i}\times(1 + {\cal A}_i)/2$ and
$n_{i2} = n_{i}\times(1 - {\cal A}_i)/2$.
The event yields $n_i$ and asymmetries ${\cal A}_i$
in each category are obtained by maximizing ${\cal L}$.
Statistical errors correspond to unit changes in the quantity 
$\chi^2 = -2\ln{({\cal L}/{\cal L}_{\mathrm{max}})}$.
The significance of a signal is measured by the square root of the change in 
$\chi^2$ when the number of signal events is constrained to zero in the 
likelihood fit;
it describes the probability for the background to fluctuate to 
the observed event yield.

The probability ${\cal P}_i(\vec{x}_j;\vec{\alpha})$ 
for a given event $j$ is the product of independent
PDFs in each of the fit input variables $\vec{x}_j$.
These variables are \DeltaE, \mes, 
\mKK, the Fisher discriminant $\cal F$, and the cosine of the $\vphi$ helicity 
angle (defined as the angle between the \Kp and $B$ momenta in the \vphi rest frame) \cosThH. 
In addition, in the simultaneous fit for the modes \phiKpm and \phipipm
we include normalized residuals 
derived from the difference between the measured and expected 
DIRC Cherenkov angles for the charged 
primary daughter.
Additional separation between the two final states is provided by 
\DeltaE. The \DeltaE separation depends on the momentum of the 
charged primary daughter in the laboratory frame and is about 
45 \mev on average, varying from about 30 \mev for the highest-momentum 
to about 80 \mev for the lowest-momentum 
primary daughters available in our final states.
If a given event has
multiple combinations satisfying the selection requirements
(which occurs in fewer than 0.2\% of the events), the ``best''
combination is selected using a $\chi^2$ quantity computed using all input variables
with the exception, in the \phihpm case, of the normalized Cherenkov-angle residuals and \DeltaE, which 
are used for $\vphi\pipm/\vphi\Kpm$ separation.

The fixed parameters $\vec{\alpha}$ defining the PDFs are extracted 
for signal from Monte Carlo 
simulation and for background distributions from the on-resonance sidebands in \mes and \DeltaE (Fig.~\ref{fig:phiMass}b). 
The MC resolutions and means are adjusted, when necessary, by comparing data and simulation 
in abundant calibration channels with kinematics and topologies similar to signal,
\piDzbarKpi and \piDmKzpi. 
The PDFs for the Cherenkov-angle residuals are 
determined from samples of \DzKpi 
originating from \Dstar decays.  

We employ a double Gaussian to parametrize
the signal \DeltaE and \mes PDFs. For the background, 
a first-degree polynomial is used for \DeltaE and an
empirical phase-space function~\cite{ref:argus} is used 
for \mes. The Fisher discriminant distributions both in signal
and in background are parametrized by a Gaussian with different widths
above and below the mean. The $\vphi$-resonance shape
in signal and the real-$\vphi$ component of the continuum background
are parametrized by the relativistic spin-1 Breit--Wigner function~\cite{phi_shape:Jackson}
with the Blatt--Weisskopf damping factor correction~\cite{phi_shape:VonHippel} convoluted with 
a Gaussian resolution function ($\sigma=1.0\mevcc$); the combinatorial component of the
\mKK distribution in the continuum background is parametrized with
a second-degree polynomial (Fig.~\ref{fig:phiMass}a).
Since \phiK and \phipi are decays of a pseudoscalar particle into a vector and a pseudoscalar, the helicity-angle 
distribution for the signal is $\cos^2\theta_H$; the background shape is 
again separated into contributions 
from combinatorial sources and from real $\vphi$ mesons, both of which are
parametrized by second-degree polynomials with no linear terms.  
The Cherenkov-angle--residual PDFs are unit Gaussians
for both the pion and kaon distributions.

For all modes, we test the fitting procedure
with background samples generated according to the PDFs and signal from Monte Carlo simulation, 
with numbers of signal and background events close to the expected. 
Signal yields are found to be unbiased.
Correlations among the input variables in data are less than 5\%.

\section{Physics Results and Systematic Uncertainties}
\label{sec:Physics}

\begin{figure}[p]
\begin{center}
\begin{tabular}{c c}
  \includegraphics[width=2.25in,height=1.95in]{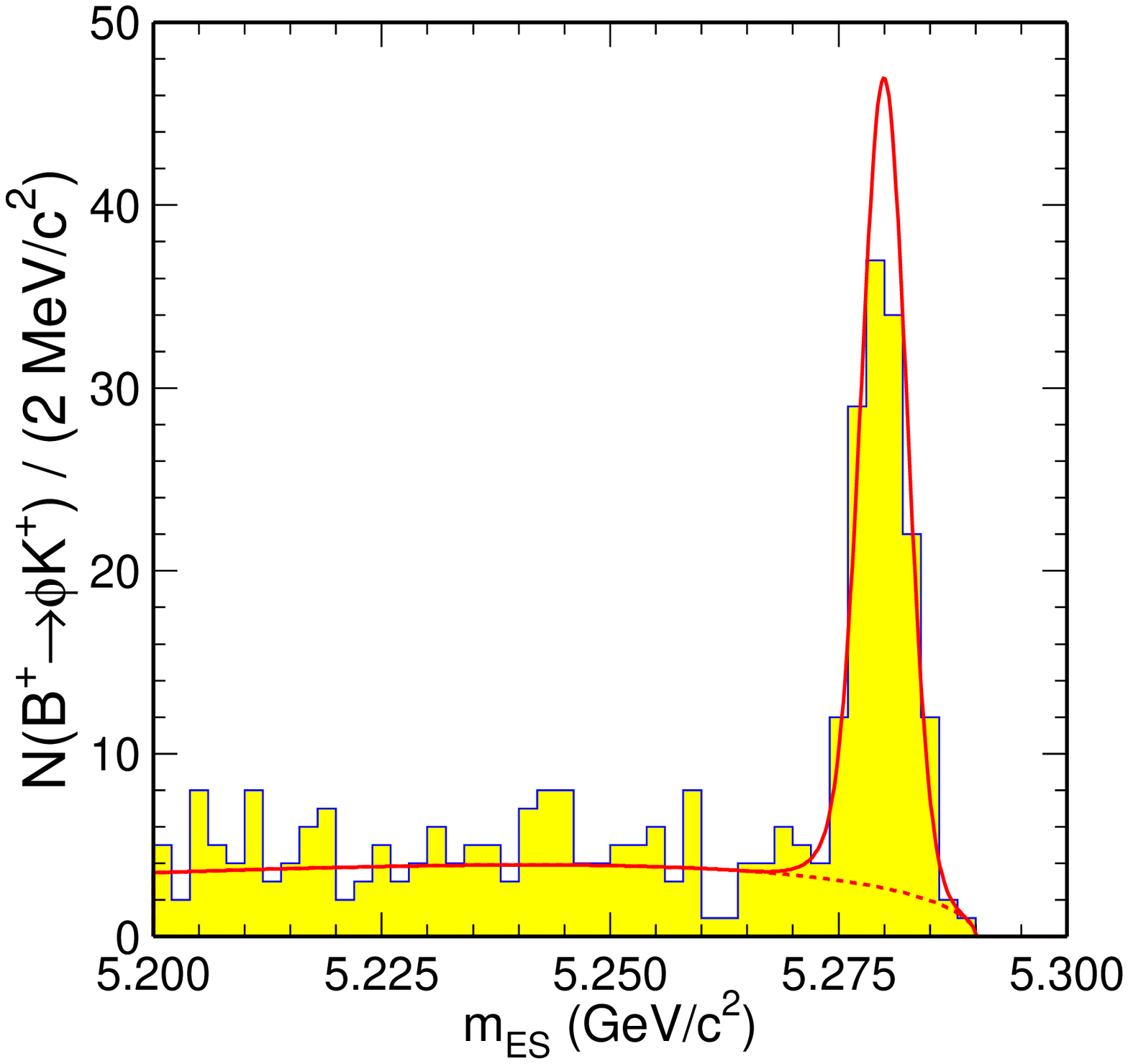} & \includegraphics[width=2.25in,height=1.95in]{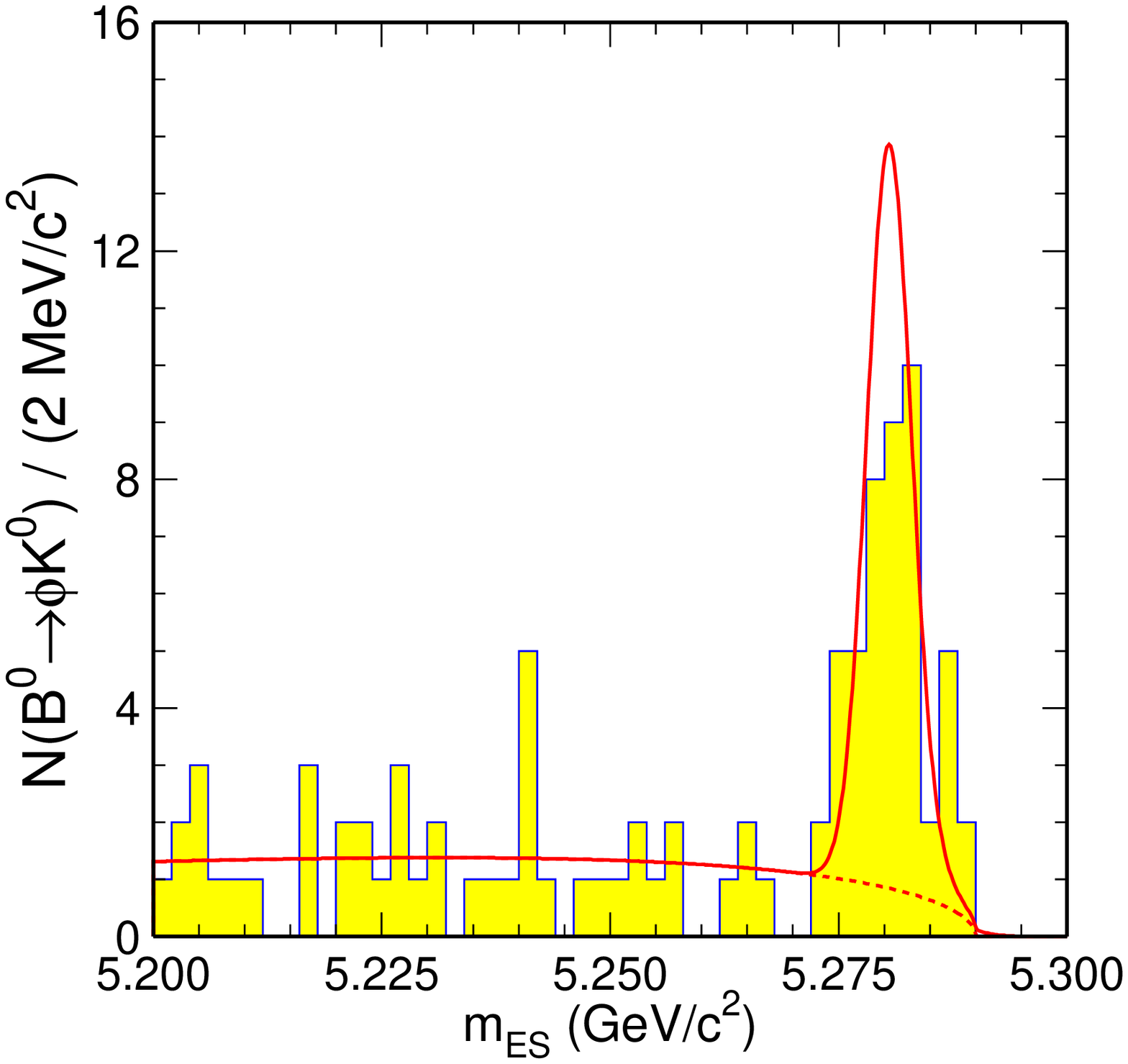} \\  
  \includegraphics[width=2.25in,height=1.95in]{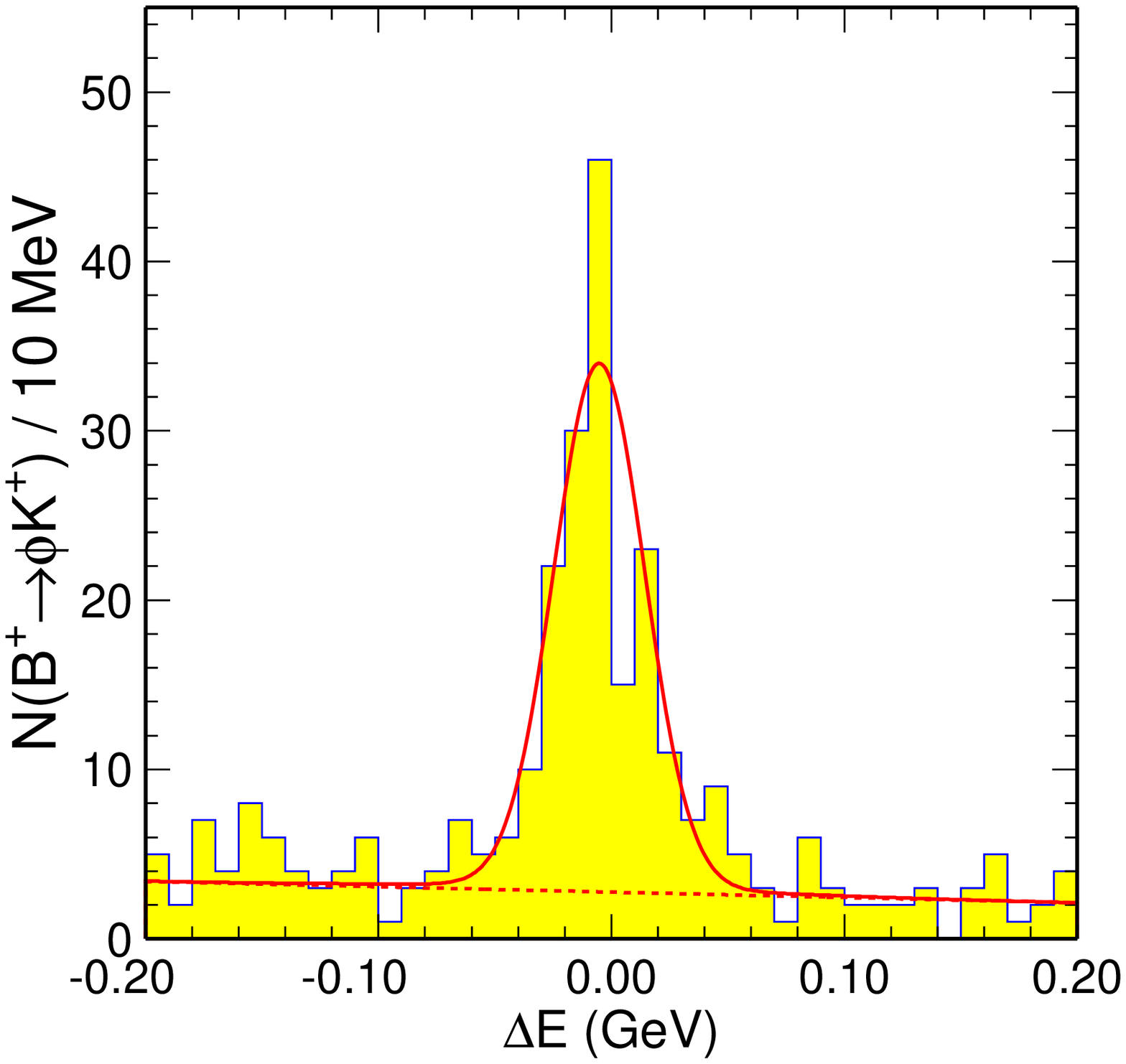} & \includegraphics[width=2.25in,height=1.95in]{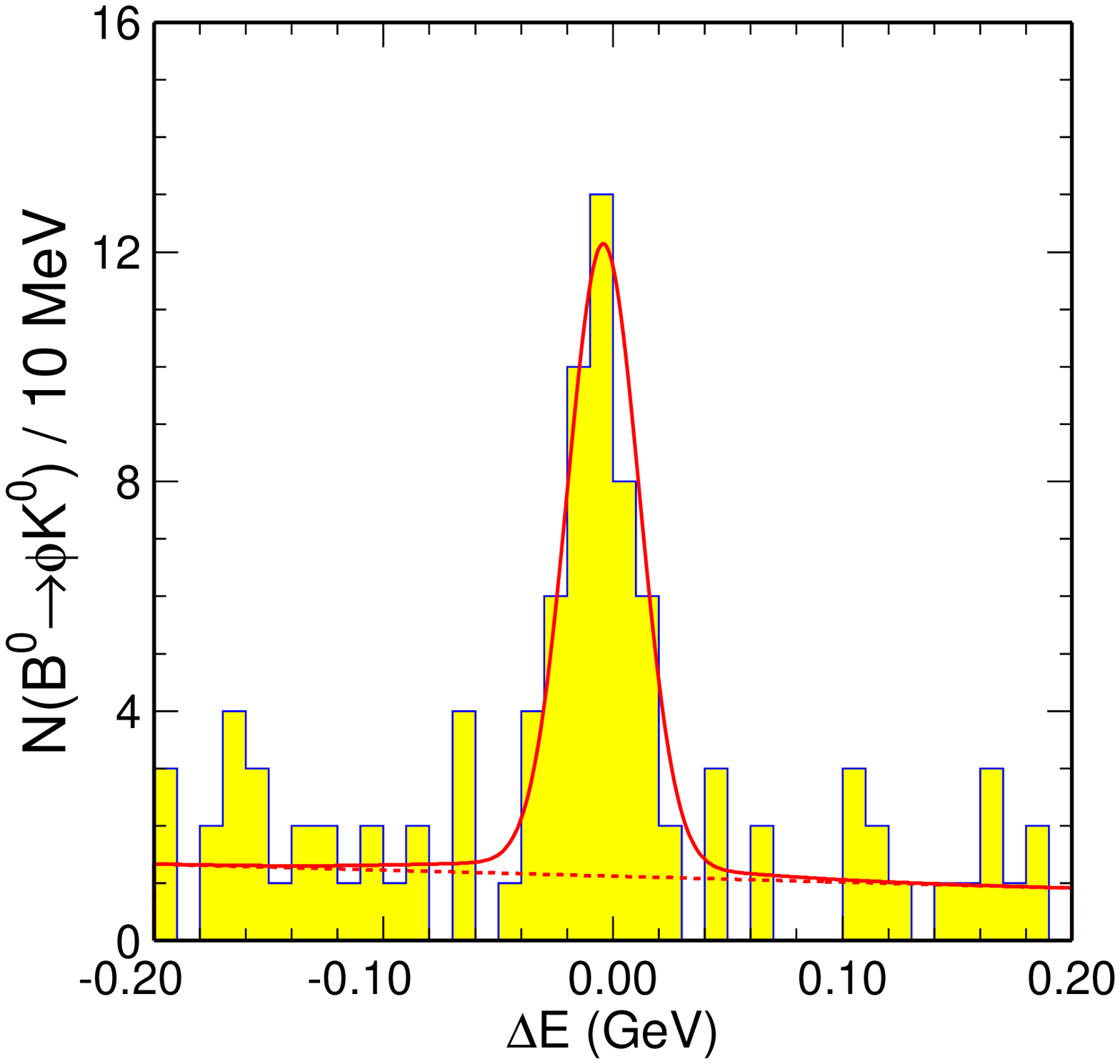} \\  
  \includegraphics[width=2.25in,height=1.95in]{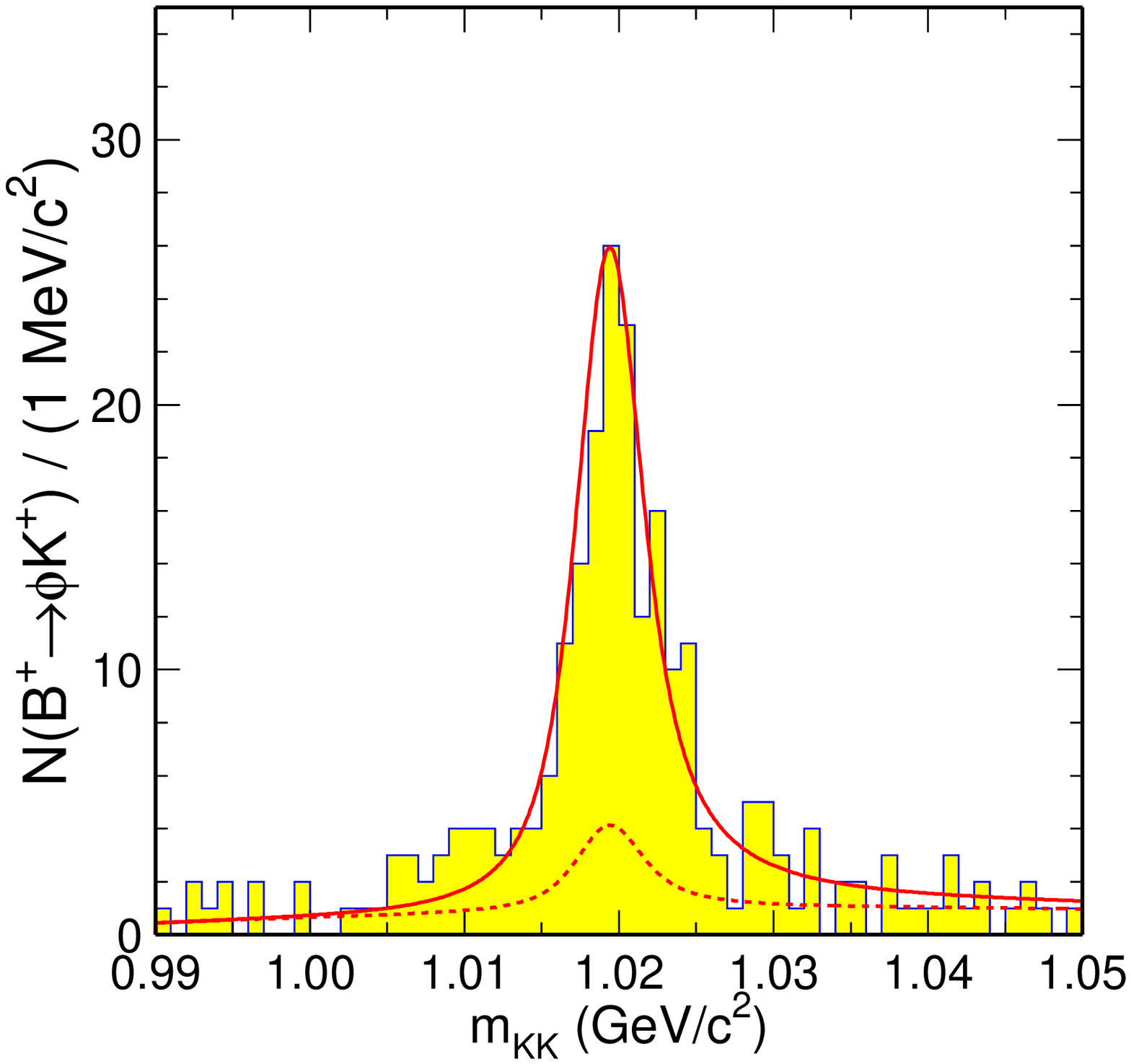} & \includegraphics[width=2.25in,height=1.95in]{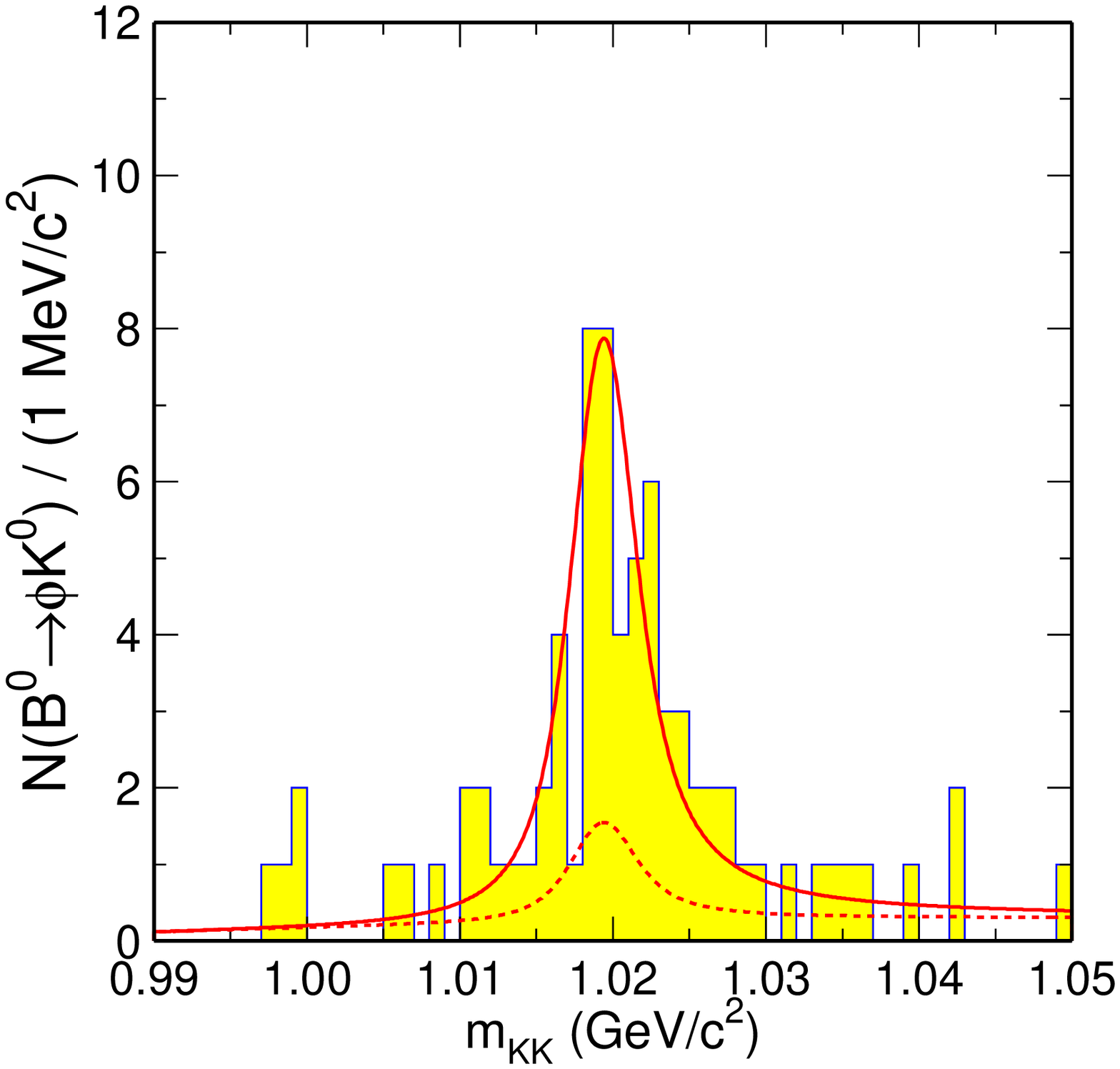} \\  
  \includegraphics[width=2.25in,height=1.95in]{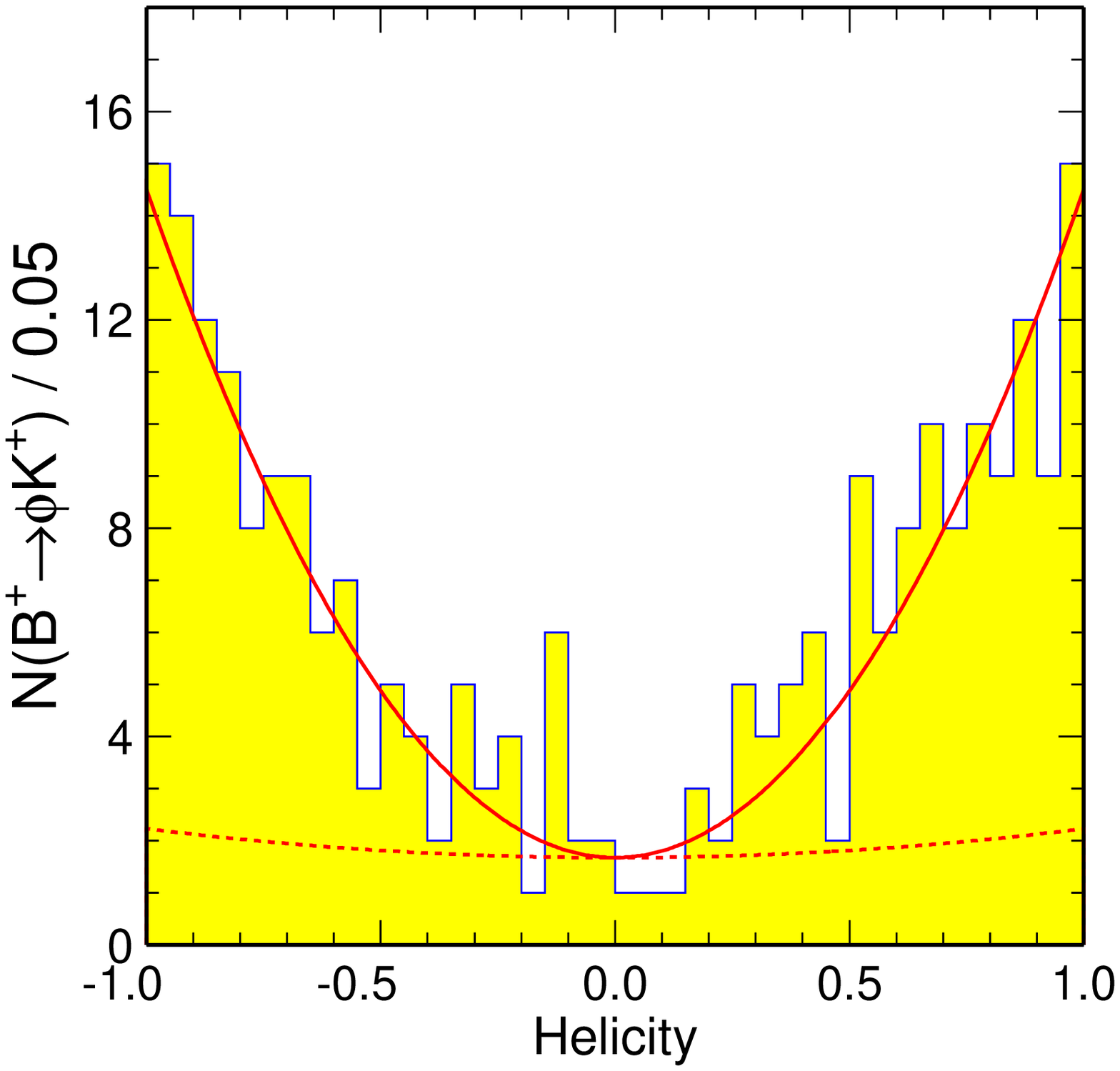} & \includegraphics[width=2.25in,height=1.95in]{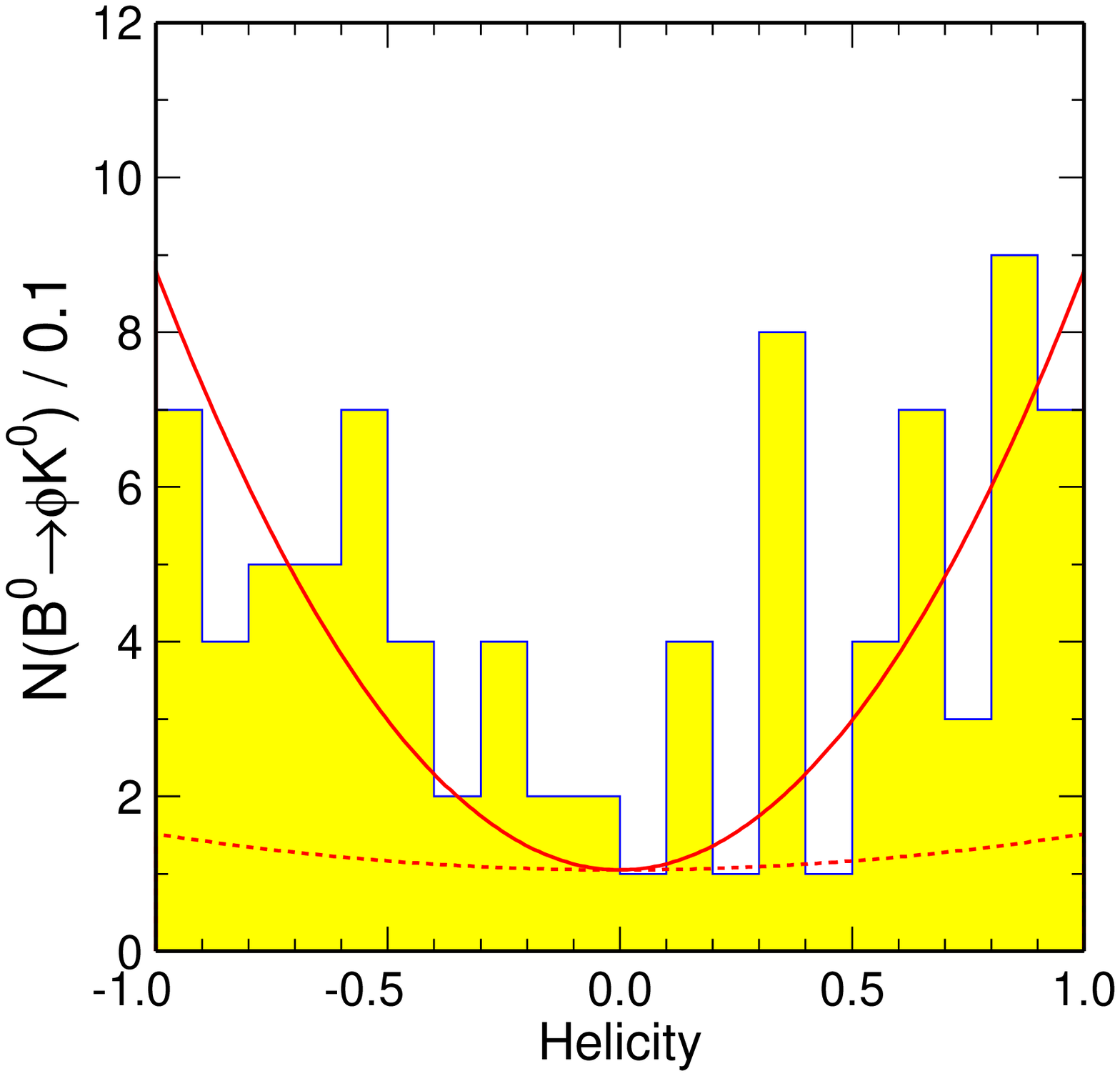} \\  
\end{tabular} 
  \put(-313,273){\small \bfseries (a)}
  \put(-138,273){\small \bfseries (b)}
  \put(-313,129){\small \bfseries (c)}
  \put(-138,129){\small \bfseries (d)}
  \put(-313,-17){\small \bfseries (e)}
  \put(-138,-17){\small \bfseries (f)}
  \put(-313,-161){\small \bfseries (g)}
  \put(-138,-161){\small \bfseries (h)}
  \caption{\small Projection plots for \phiKpm (left column) and \phiKz (right column), 
    made with a probability-ratio cut to emphasize the signal, for the variables \emph{(a,b)}~\mes, 
    \emph{(c,d)}~\DeltaE, \emph{(e,f)}~\mKK, \emph{(g,h)}~\cosThH.
    The solid (dashed) lines show the signal+background (background only) PDF projections}
  \label{fig:projections}
\end{center}
\end{figure}

The results of our ML fit analyses are summarized in Table~\ref{table:results}.
For the branching fractions, equal production rates of 
\BzBzb and \BpBm are assumed.
The projections of the fit results are shown in Fig.~\ref{fig:projections}, where 
we plot only a subsample of events, enhancing the signal with a requirement 
on the ratio of probabilities for each event to belong either to the signal or to the
background categories.

\begin{table}[!tb]
\begin{center}
\caption{\small Summary of results. Equal production rates of 
\BzBzb and \BpBm are assumed. The total efficiency values include daughter branching fractions. Central values are followed
by statistical and systematic errors; the upper limit on \BR(\phipipm) incorporates the associated systematic error. The 
statistical significance of the \phipipm signal is $0.5\sigma$. The 90\% confidence interval for \ACP(\phiKpm) is 
$[-0.104;+0.181]$}
\vspace*{0.1in}
\begin{tabular}{|c|c|c||c|}
\hline
\hline
                                  & \phiKpm \high3pad       &  \phipipm \low4pad              &     \phiKz        \\ 
\hline  
Events to fit                     & \multicolumn{2}{c||}{14371~~~~~~\,}                        &  2043              \\ 
Signal yield                      & $173 \pm 15$ & $0.9^{+2.8}_{-0.9}$  ($<6.7$ at 90\% CL) &  $50^{+9}_{-8}$ \low5pad \\
\hline
Reconstruction eff.~(\%)                    &  39.8                  &   41.4                 & 43.5 \\
Total efficiency (\%)                   &  19.6                  &   20.4                 & 7.4             \\
\hline
\BR\ ($10^{-6}$)                   & $10.0^{+0.9}_{-0.8}\pm0.5$ & $<0.41$ (90\% CL)       & $7.6^{+1.3}_{-1.2}\pm0.5$ \high4pad \low5pad \\
            \ACP                  & $0.039 \pm 0.086 \pm 0.011$ &  ---  &  --- \low4pad \\ 
\hline
\hline
\end{tabular}
\label{table:results}
\end{center}
\end{table} 

\label{sec:Systematics}

Systematic uncertainties in the ML fit originate from assumptions about
the signal and background distributions.
We simultaneously vary all PDF parameters within their uncertainties
and derive the associated systematic errors,
which are found to be 2.0\% for $\BR(\phiKpm)$, 10.9\% for the 90\% upper limit on
$\BR(\phipipm)$, 2.8\% for $\BR(\phiKz)$, and 0.5\% for $\ACP(\phiKpm)$.

The dominant systematic errors in the efficiency come from track 
finding (0.8\% per high-momentum or \phiKK track),
particle identification (1\% per \phiKK track), 
and \KS
reconstruction efficiency (4.0\%). 
Other minor systematic effects
from event-selection criteria, daughter branching fractions,
MC statistics, \BB backgrounds and $B$-meson counting sum to 3.0\%.
Efficiency uncertainties affect the values of the branching 
fractions, but not their significances. The systematic uncertainty
on \ACP due to charge asymmetries in tracking and the DIRC is less than 1.0\%.

Given the low significance of $\BR(\phipipm)$,
we quote a 90\% CL upper limit obtained by integrating the normalized 
likelihood distribution from zero.
The limit incorporates changes by one standard deviation
from uncertainties in PDFs and the reconstruction efficiency.

\section{Summary}
\label{sec:Summary}

We have determined the branching fractions 
of the rare charmless penguin-dominated $B$-meson decays
\phiKpm and \phiKz, and have set a limit on the direct \emph{CP} asymmetry $\ACP(\phiKpm)$, with 
substantially reduced statistical and systematic errors compared to previously
published results. The results contained in this paper are preliminary.
 
The stringent upper limit on the CKM-- and color-suppressed decay \phipipm
provides evidence against the presence of large non-penguin or non--Standard Model contributions
to the $b\to s(d)\sbar \s$ decay amplitudes. 

\section{Acknowledgements}
\label{sec:Acknowledgments}

We are grateful for the 
extraordinary contributions of our \pep2\ colleagues in
achieving the excellent luminosity and machine conditions
that have made this work possible.
The success of this project also relies critically on the 
expertise and dedication of the computing organizations that 
support \babar.
The collaborating institutions wish to thank 
SLAC for its support and the kind hospitality extended to them. 
This work is supported by the
US Department of Energy
and National Science Foundation, the
Natural Sciences and Engineering Research Council (Canada),
Institute of High Energy Physics (China), the
Commissariat \`a l'Energie Atomique and
Institut National de Physique Nucl\'eaire et de Physique des Particules
(France), the
Bundesministerium f\"ur Bildung und Forschung and
Deutsche Forschungsgemeinschaft
(Germany), the
Istituto Nazionale di Fisica Nucleare (Italy),
the Foundation for Fundamental Research on Matter (the Netherlands),
the Research Council of Norway, the
Ministry of Science and Technology of the Russian Federation, and the
Particle Physics and Astronomy Research Council (United Kingdom). 
Individuals have received support from 
the A. P. Sloan Foundation, 
the Research Corporation,
and the Alexander von Humboldt Foundation.



\begin{thebibliography}{99}

\bibitem{ref:theory1}
N.G.~Deshpande and J.~Trampetic,
Phys.\ Rev.\ D {\bf 41}, 895 (1990);
N.G.~Deshpande and X.-G.~He,
Phys.\ Lett.\ B {\bf 336}, 471 (1994);
R.~Fleischer,
Z.\ Phys.\ C {\bf 62}, 81 (1994).

\bibitem{ref:newphys}
I.~Hinchliffe and N.~Kersting,
Phys.\ Rev.\ D {\bf 63}, 015003 (2001).

\bibitem{ref:phiKSexp_BaBar}
B.~Aubert {\it et al.}  [BABAR Collaboration],
arXiv:hep-ex/0207070. 

\bibitem{ref:phiKSexp_Belle}
K.~Abe {\it et al.}  [Belle Collaboration],
arXiv:hep-ex/0207098.

\bibitem{Grossman:1997gr}
Y.~Grossman, G.~Isidori and M.~P.~Worah,
Phys.\ Rev.\ D {\bf 58}, 057504 (1998).

\bibitem{Cheng:2000hv}
H.~Y.~Cheng and K.~C.~Yang,
Phys.\ Rev.\ D {\bf 64}, 074004 (2001).

\bibitem{Chen:2001pr}
C.~H.~Chen, Y.~Y.~Keum and H.~n.~Li,
Phys.\ Rev.\ D {\bf 64}, 112002 (2001).

\bibitem{Briere:2001ue}
R.~A.~Briere {\it et al.}  [CLEO Collaboration],
Phys.\ Rev.\ Lett.\  {\bf 86}, 3718 (2001).

\bibitem{Aubert:2001zd}
B.~Aubert {\it et al.}  [BABAR Collaboration],
Phys.\ Rev.\ Lett.\  {\bf 87}, 151801 (2001).

\bibitem{Bozek:2001xd}
A.~Bozek  [Belle Collaboration],
arXiv:hep-ex/0104041.


\bibitem{ref:babar}
B.~Aubert {\it et al.}  [BABAR Collaboration],
Nucl.\ Instrum.\ Meth.\ A {\bf 479}, 1--116 (2002).

\bibitem{ref:pep} 
PEP-II Conceptual Design Report, SLAC-R-418 (1993).

\bibitem{Fisher:et}
R.~A.~Fisher,
Annals Eugen.\  {\bf 7}, 179 (1936).

\bibitem{ref:argus}
H.~Albrecht {\it et al.} [ARGUS Collaboration], Phys.\ Lett.\ B {\bf 241}, 278 (1990).

\bibitem{phi_shape:Jackson}
J.~D.~Jackson,
Nuovo Cim.\  {\bf 34}, 1644 (1964).

\bibitem{phi_shape:VonHippel}
F.~Von Hippel and C.~Quigg,
Phys.\ Rev.\ D {\bf 5}, 624 (1972).

\end{thebibliography}
\end{document}